# ENZ materials and Anisotropy: Enhancing Nonlinear Optical Interactions at the Nanoscale in Metal/Conducting-Oxide Multilayer Stacks

M. A. Vincenti[1], D. de Ceglia[2], and M. Scalora[3]

[1]Department of Information Engineering- University of Brescia, Via Branze 38, 25123 Brescia, Italy
[2]Department of Information Engineering – University of Padova, Via Gradenigo 6/A, 35133 Padova, Italy
[3]Charles M. Bowden Laboratory, CCDC-AvMC, Redstone Arsenal, AL 35898-5000, USA
*Corresponding author: maria.vincenti@unibs.it

## Abstract

Epsilon-near-zero materials are exceptional candidates for studying electrodynamics and nonlinear optical processes at the nanoscale. We demonstrate that by alternating a metal and a highly doped conducting-oxide, the epsilon-near-zero regime may be accessed resulting in an anisotropic, composite nanostructure that significantly enhances nonlinear interactions. Using two independent and different computation techniques we show that the structure can enhance the local field intensity by nearly two orders of magnitudes, in large part due to the onset of the effective anisotropy. The investigation of the multilayer nanostructure using a microscopic, hydrodynamic approach also sheds light on the roles of two competing contributions that are for the most part overlooked, but that can significantly modify linear and nonlinear responses of the structure: nonlocal effects, which blueshift the resulting resonance, and the hot electron nonlinearity, which redshifts the plasma frequency as the effective mass of free electrons increases as a function of incident power density. Finally, we show that, even in absence of second order bulk nonlinearity, second order nonlinear processes are also significantly enhanced by the layered structure.

## Introduction

In the last few decades the scientific community has focused significant effort on the design and fabrication of artificial nanostructures, with the intent of circumventing limitations imposed by natural materials, like absorption, and provide novel functionalities by harnessing processes that may occur only at the nanoscale, while simultaneously reducing the size of optical devices. Among the plethora of proposed novel materials and nanostructures, those operating near their epsilon-near-zero (ENZ) region have been shown to possess a set of peculiar properties [1-3]. These materials may be exploited to manipulate radiation direction and polarization [4-8], but may also serve as platforms to study nonlinear optical interactions thanks to their ability to enhance local fields under specific circumstances, i.e., TM-polarized field incident at oblique incidence [3]. Among the explored nonlinear processes [9-16] that have been studied both theoretically and experimentally [17-26], second and third harmonic generation stand out.. The ENZ condition is known to manifest itself in any natural material. For example, GaAs, Si, and GaP have their zero crossing for the real part of the permittivity in the UV, metals like Au, Ag and Cu in the visible, and highly-doped oxides like ITO and AZO in the infrared [27]. However, it is possible to design artificial nanostructures and tailor the frequency response at will [28] or to compensate losses [18, 29, 30], which are thought to represent the main limitation for field enhancement in ENZ materials [31]. Still, loss compensation techniques based on the inclusion of active materials might be challenging for fabrication and not practical for certain applications.

An alternative route to achieving high local field enhancement without resorting to loss compensation techniques is to exploit the effective anisotropy of artificial nanostructures via the realization of longitudinal epsilon-near-zero, or LENZ, materials [32]. Indeed, it has been shown that LENZ possess exceptional abilities that span from perfect light bending [33, 34] and angular filtering and polarization control [35], to coherent perfect absorption [36] and control of leaky wave radiation [37]. Moreover, since the effective anisotropy tends to enhance the local field intensity and improve tolerances in terms of acceptance angles and a weaker sensitivity to material thickness [32], LENZ turn out to be good candidates for the investigation of nonlinear processes. Recently, LENZ have been studied in the context of second order nonlinear optical processes arising from symmetry breaking at each interface of a layered structure. It was shown that they can boost the enhancement of second harmonic generation (SHG) beyond their isotropic counterparts [38]. Here we first investigate third harmonic generation (THG) with the intent to further examine the role of anisotropy on nonlinear processes in ENZ media. We simulate an effective anisotropic medium and demonstrate that, regardless of the thickness of the medium, THG can be enhanced dramatically by properly increasing the material's degree of anisotropy. We then analyze a realistic scenario, where a metal/conducting-oxide multilayer stack composed of Au and Dy:CdO (dysprosium-doped Cadmium Oxide) is used to test the impact of anisotropy on the same third order process. We compare the response of the multilayer with and without the effects of nonlocality, and introduce the contribution of hot electrons, two processes that can be triggered in both metals and conducting oxides, and compete to either blueshift or redshift the resonant linear and nonlinear responses of the structure. Finally, we demonstrate that the high degree of anisotropy achieved in a multilayer environment also enhances second order nonlinear processes even in absence of bulk nonlinearities.

**Improving nonlinear phenomena through anisotropy**

Anisotropic ENZ or LENZ have been shown to increase SHG conversion efficiencies in the absence of dipole-allowed quadratic nonlinearity [38]. This improvement is possible thanks to the high local fields that can be achieved in anisotropic ENZ [32] even without the aid of geometric resonances. Although it is clear from previous investigations that anisotropy facilitates nonlinear optical processes in the presence of the ENZ condition, to date it is still not clear if nanostructures with a low degree of anisotropy can perform as well as structures with a high degree of anisotropy, and whether or not the improvement associated with the anisotropy also applies to bulk nonlinearities.

In order to shed light on these two aspects we first investigate THG in a homogenous slab of material having the dielectric permittivity of Dy:CdO [39]. Then we artificially modify the transverse components of the dielectric permittivity ($\varepsilon_z = \varepsilon_{Dy:CdO}$, $\varepsilon_x = \varepsilon_y = \varepsilon_{Dy:CdO} + \Delta\varepsilon$) to simulate a Dy:CdO-like material and introduce an arbitrary degree of anisotropy. For simplicity, in this section we assume the material is isotropic at the third harmonic frequency. The geometry of the structure under consideration is depicted in Fig.1: a slab of thickness $d$ is illuminated by a TM-polarized field incident at a variable angle $\vartheta_i$.

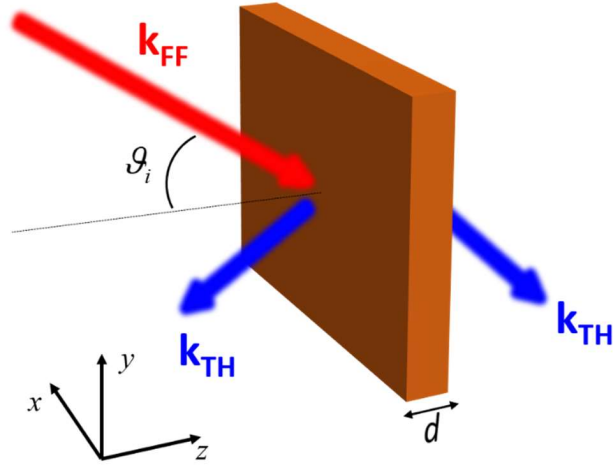

**Fig.1:** Sketch of the structure under investigation: a slab of material of thickness *d* is illuminated by a TM-polarized pump wavevector **k**$_{FF}$ at an angle $\vartheta_i$. Forward and backward third harmonic signals are generated with wavevector ±**k**$_{TH}$, respectively.

The fundamental frequency (FF) corresponds to $\lambda_{FF} = 2147$ nm, where the real part of the dielectric permittivity of Dy:CdO displays a zero crossing. Pump irradiance is set to 1 GW/cm². In this preliminary analysis we solve both linear and nonlinear electromagnetic problems by using a frequency-domain, finite-element solver (COMSOL Multiphysics). Since we are simulating a highly-doped oxide, we assume the material possesses an isotropic, bulk, dispersion-less and near-instantaneous $\chi^{(3)} = 10^{-20}$ m²/V² at both FF and third harmonic (TH) frequency. This value is compatible with those measured for other doped oxides, such as ITO and AZO [14, 17]. No dipole-allowed second-order nonlinearity is considered. The TM-polarized field is assumed to be a superposition of two monochromatic signals at the FF and TH harmonic frequencies. The TH signal is initially null. Therefore, the time-independent, nonlinear complex current densities at FF and TH may be written as [40]:

$$\mathbf{J}_{FF}^{NL} = -i3\varepsilon_0 \omega_{FF} \chi^{(3)} \left( \mathbf{E}_{FF} \cdot \mathbf{E}_{FF}^* \right) \mathbf{E}_{FF}, \qquad (1)$$

$$\mathbf{J}_{TH}^{NL} = -i\varepsilon_0 \omega_{TH} \chi^{(3)} \left( \mathbf{E}_{FF} \cdot \mathbf{E}_{FF} \right) \mathbf{E}_{FF}. \qquad (2)$$

In order to understand the impact of anisotropy we begin our investigation by calculating the field intensity enhancement, FIE $= |\langle E_z \rangle / E_0|^2$, where $\langle E_z \rangle$ is the average value of the component of the electric field in the direction of propagation (see Fig.1), evaluated inside the slab. $E_0$ is the amplitude of the incident electric field. In Fig. 2 we show FIE maps for different slab thicknesses, respectively for $d = 20$ nm [Fig. 2(a)], $d = 60$ nm [Fig. 2(c)], and $d = 100$ nm [Fig. 2(e)]. FIE is calculated as a function of the angle of incidence and degree of anisotropy $\Delta\varepsilon$. Specifically, the larger the value of $|\Delta\varepsilon|$ is, the larger the difference will be between longitudinal (direction of propagation) and transverse permittivities.

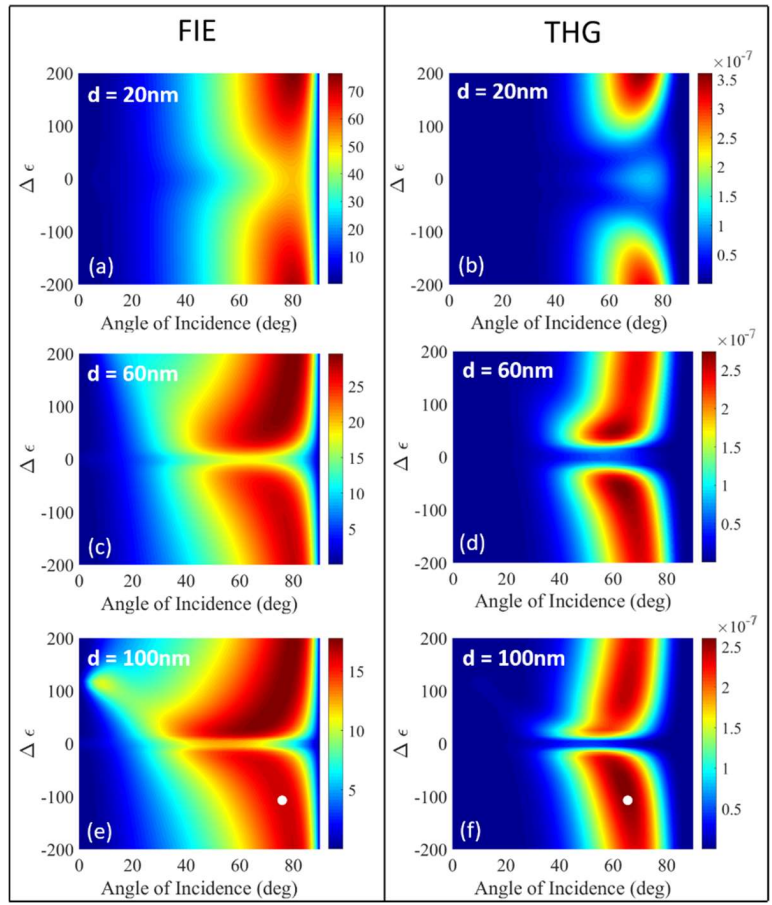

**Fig.2** FIE for a slab of material with variable anisotropy and angle of incidence for thicknesses (a) $d = 20$ nm, (c) $d = 60$ nm, and (e) $d = 100$ nm; Total third harmonic generated signal (forward plus backward) for a slab of material with variable anisotropy and angle of incidence and thickness (b) $d = 20$ nm, (d) $d = 60$ nm, and (f) $d = 100$ nm.

As one departs from the isotropic condition ($\Delta\varepsilon = 0$), FIE increases for a certain range of angles of incidence, regardless of slab thickness. We note that the sign of the anisotropy also appears to impact FIE, which attains slightly larger values when $\Delta\varepsilon$ is positive. The plots also confirm that the tolerance to the angle of incidence improves as the thickness of the slab is increased, as indicated in Ref. [32]. Although FIE improvement is generally a good guideline when attempting to boost nonlinear optical processes, the maps of total (forward plus backward) THG for thicker slabs [Figs. 2 (d) and (f)] show a somewhat different trend: while Figs. 2 (c) and (e), show that FIE reaches higher values when $\Delta\varepsilon \gg 0$, the THG maps in Figs. 2 (d) and (f) reveal that total third harmonic conversion efficiency generally improves when the degree of anisotropy increases, but achieves larger values when $\Delta\varepsilon \ll 0$. The reason behind the discrepancy between the FIE and THG trends in thicker slabs resides in the definition of FIE, which considers the average value of the longitudinal electric field inside the slab. In fact, the maximum value of the longitudinal electric field component $E_z$ (not shown), which occurs at the first interface between vacuum and the slab, is largest for large negative anisotropy ($\Delta\varepsilon \ll 0$), and produces a stronger TH signal. Put another way, while a large degree of anisotropy is desirable in order to improve FIE, the sign of the anisotropy may also be relevant to determine the efficiency of the nonlinear optical process, especially in thicker slabs.

**Third-Harmonic Generation from Au/Dy:CdO multilayer stack**

A careful look at Fig. 2 reveals that relatively high |Δε| are generally needed to significantly improve the efficiency of nonlinear processes compared to the isotropic case (Δε = 0). For example, a 100nm thick slab shows a maximum THG efficiency occurring near 65°, when Δε ≅ -100. At first sight, the problem of assembling a practical device may appear to be daunting. The requirement is to achieve high anisotropy while preserving a zero-crossing in the real part of the effective permittivity in the longitudinal direction. However, if we aim to realize an anisotropic material with $\Delta\varepsilon \ll 0$, one may resort to the inclusion of metals because they naturally display large, negative permittivities in practically all spectral ranges of interest. For this reason we direct our attention to structures similar to that depicted in Fig.3: a five periods metal-dielectric stack composed of Au [27] and Dy:CdO [39], each 10 nm thick ($a = b = 10$ nm), for a total thickness $d$ = 100 nm, comparable in size to the slab of Fig. 2 (c) and (e). The multilayer is situated on top of a SiO$_2$ semi-infinite substrate, and it is illuminated by a TM-polarized field incident at a variable angle $\vartheta_i$.

The effective permittivities of the stack calculated at $\lambda_{FF}$ = 2147 nm (the crossing point of Dy:CdO) are $\varepsilon_{z,FF}$ = *2.1·10$^{-7}$ + 0.34i, $\varepsilon_{x,FF}$ = $\varepsilon_{y,FF}$ = -101.34 + 11.34i* at the FF, and $\varepsilon_{z,TH}$ = *12.46 + 0.51i, $\varepsilon_{x,TH}$ = $\varepsilon_{y,TH}$ = -6.52 + 0.92i* at the TH. While in the previous section the eventual anisotropy at the TH frequency was ignored, a realistic nanostructure will indeed display anisotropic behavior also at the TH. To boot, in our preliminary analysis we also assumed a purely real Δε. The introduction of metals in the multilayer stack adds a realistic degree of damping in the system, as revealed by the imaginary part of the effective permittivity in the transverse direction.

Metals like gold and silver typically require one free and two bound electron contributions in order to accurately reproduce the local dielectric constant over a broad range that includes UV wavelengths, where interband transitions significantly contribute to the dielectric response. Conducting oxides, on the other hand, typically require one Drude and one Lorentz electron species. Therefore, gold data were modeled by using a combination of a Drude ($\omega_{p,f} = 1.27 \cdot 10^{16}$ s$^{-1}$, $\gamma_f = 9.43 \cdot 10^{13}$ s$^{-1}$) and two Lorentz oscillators ($\omega_{p,b1} = 7.73 \cdot 10^{15}$ s$^{-1}$, $\omega_{0,b1} = 5.18 \cdot 10^{15}$ s$^{-1}$, $\gamma_{b1} = 2.26 \cdot 10^{15}$ s$^{-1}$, $\omega_{p,b2} = 1.12 \cdot 10^{16}$ s$^{-1}$, $\omega_{0,b2} = 8.48 \cdot 10^{15}$ s$^{-1}$, $\gamma_{b2} = 2.26 \cdot 10^{15}$ s$^{-1}$) that provide a good fit for data down to approximately 200nm [27]. On the other hand, Dy:CdO [39] data were fit with a single Drude oscillator ($\omega_{p,f} = 1.99 \cdot 10^{15}$ s$^{-1}$, $\gamma_f = 2.73 \cdot 10^{13}$ s$^{-1}$) for free electrons, and one Lorentz oscillator for bound electrons with a resonance far from the ENZ condition ($\omega_{p,b1} = 8.87 \cdot 10^{16}$ s$^{-1}$, $\omega_{0,b1} = 4.33 \cdot 10^{16}$ s$^{-1}$, $\gamma_{b1} = 1.88 \cdot 10^{15}$ s$^{-1}$). As it was done for the single slab, we evaluate both linear and nonlinear responses of the structure so that we may keep track of how the two regimes are related. In order to model a realistic metal-dielectric stack we resort to a detailed, microscopic hydrodynamic description of light-matter interactions that has previously been used to simulate wave propagation in an ITO nanolayer [41-44]. We also simulated both linear and nonlinear electromagnetic problems by means of a frequency-domain, finite-element solver (COMSOL Multiphysics). This will help us verify that the results are robust and consistent in the continuous wave and pulsed regimes. Using the finite element solver, gold and Dy:CdO layers are modeled with a near-instantaneous, isotropic nonlinear response with $\chi^{(3)} = 10^{-20}$ m$^2$/V$^2$. On the other hand, the hydrodynamic approach allows for a realistic description of both linear and nonlinear dispersions without a preconceived notion of the magnitude of $\chi^{(3)}$. Both approaches yield qualitatively similar results. We stress that, while gold has been reported to have a nonlinear susceptibility of the order of 10$^{-16}$ m$^2$/V$^2$ in the visible range, its magnitude generally depends on pulse duration and may decrease substantially away from the plasma frequency of the material [14], a fact that can easily be verified using the Drude-Lorentz oscillator approach. In addition, the fields are strongly localized inside the CdO layers, a circumstance that greatly diminishes the nonlinear response of gold.

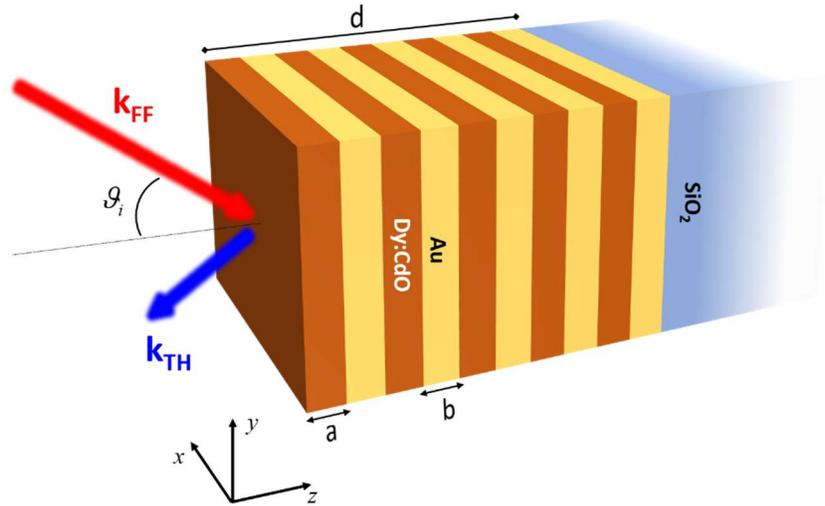

**Fig.3** Sketch of the multilayer structure: five periods of Dy:CdO (a=10nm) and Au (b=10nm) are alternated to obtain an anisotropic response. The structure is illuminated by a TM-polarized pump wavevector $\mathbf{k}_{FF}$ at an angle $\vartheta_i$. Only backward third harmonic signal with wavevector $\mathbf{k}_{TH}$ is monitored.

Pump irradiance is assumed once again to be 1 GW/cm². FIE and THG maps are shown in Fig. 4 as functions of incident wavelength and angle of incidence. In the multilayer $\text{FIE} = |\langle E_z \rangle / E_0|^2$ is obtained by evaluating the average value of the component of the electric field in the direction of propagation $\langle E_z \rangle$ inside all layers of the stack, keeping in mind, however, that field localization is localized mostly inside the CdO layers. Both FIE and THG peak in the vicinity of the zero-crossing for the effective permittivity in the propagation direction. The peaks are also relatively wide as a function of the angle of incidence, in contrast with the response of isotropic ENZ materials [38]. We note that the TH conversion efficiency shown in Fig. 4 (b) refers only to the TH reflected signal. The transmitted TH signal is two orders of magnitude smaller than the reflected component and is, therefore, ignored. Although the peak values shown in Fig. 4 are similar to the results presented in the previous section for the 100 nm slab having similar values of longitudinal and transverse permittivities, [see white dots in Figs. 2 (e) and (f)], we are now analyzing a nanostructure that presents a non-homogeneous electric field distribution. Therefore, the concept of effective permittivity falls short of an adequate description of nonlinear interactions and may only be used as a guideline to design multilayers. Care should be exercised when predicting the nonlinear response from the structure using effective medium approaches.

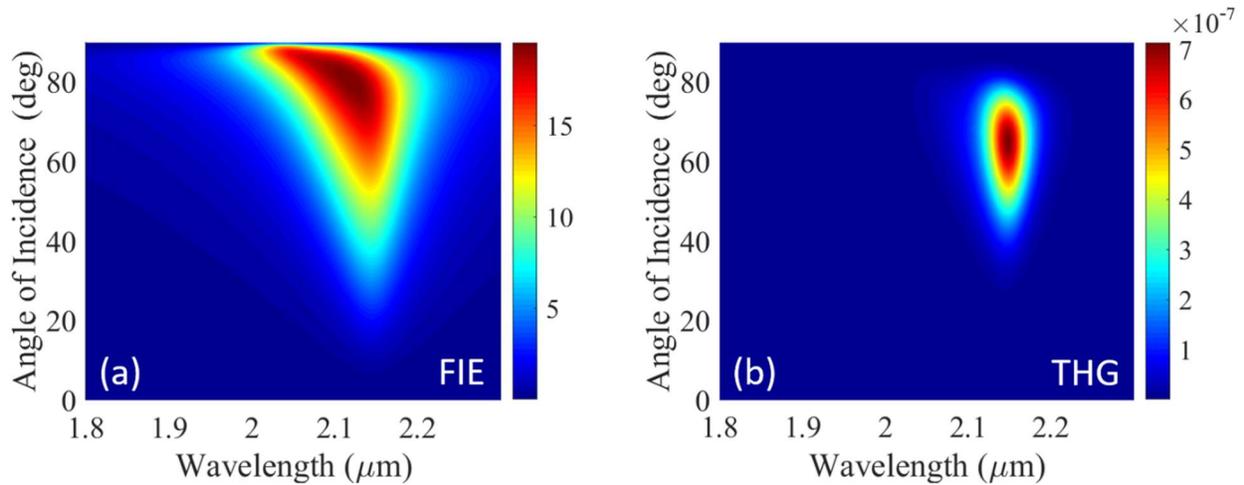

**Fig.4** (a) FIE and (b) TH conversion efficiency for the multilayer structure sketched in Fig. 3, calculated as a function of incident wavelength and angle of incidence.

In the multilayer scenario, three important aspects need to be taken into account for both linear and nonlinear simulations: (i) the thickness of each layer is 10 nm, so that nonlocal contributions should be taken into account in both linear and nonlinear regimes; (ii) it is known that hot carrier contributions may be triggered in both Au and Dy:CdO, significantly altering the linear and nonlinear responses [42]; (iii) a second-order response that arises from symmetry breaking at the interfaces in not negligible in this kind of structure [38], may lead to cascaded THG [26], and must be included in the calculations [42, 44, 45].

Therefore, in addition to the nonlinear contributions that arise from bound electrons, which peak in a different wavelength range [42] and are taken into account directly into Maxwell's equations, the nonlinearities associated with equations (1) and (2) are now related directly to hot electrons, and are treated in the Drude portion of the simulation. Adding the usual nonlocal terms to equations (1) and (2), with the understanding that we are now dealing with free electrons, we obtain the following novel equations of motion for the nonlinear, free currents:

$$\frac{3v_F^2/5}{\omega_{FF}(\omega_{FF}+i\gamma_f)}\nabla(\nabla\cdot\mathbf{J}_{FF})+\mathbf{J}_{FF}-i\omega_{FF}\chi^{(3)}_{hot,FF}\left(\mathbf{E}_{FF}\cdot\mathbf{E}^*_{FF}\right)\mathbf{E}_{FF}=\frac{\varepsilon_0\omega^2_{p,f}}{\gamma_f-i\omega_{FF}}\mathbf{E}_{FF}, \qquad (3)$$

$$\frac{3v_F^2/5}{\omega_{TH}(\omega_{TH}+i\gamma_f)}\nabla(\nabla\cdot\mathbf{J}_{TH})+\mathbf{J}_{TH}-i\omega_{TH}\chi^{(3)}_{hot,TH}\left(\mathbf{E}_{FF}\cdot\mathbf{E}_{FF}\right)\mathbf{E}_{FF}=\frac{\varepsilon_0\omega^2_{p,f}}{\gamma_f-i\omega_{TH}}\mathbf{E}_{TH}. \qquad (4)$$

where $v_F=\hbar(3\pi^2 n_0)^{1/3}/m_f^*$ is the Fermi velocity (we assumed $v_F=1.39\cdot 10^6$ m/s for Au and $v_F=1.34\cdot 10^6$ m/s for Dy:CdO); $\omega_{p,f}=\sqrt{n_0 e^2/(\varepsilon_0 m_f^*)}$ is the unscreened bulk plasma frequency ($\omega_{p,f}=1.27\cdot 10^{16}$ s$^{-1}$ for Au and $\omega_{p,f}=1.99\cdot 10^{15}$ s$^{-1}$ for Dy:CdO); $n_o$ is the (constant) equilibrium carrier density; $m_f^*$ is the effective electron mass; $\gamma_f$ is the scattering rate due to collisions ($\gamma_f=9.43\cdot 10^{13}$ s$^{-1}$ for Au and $\gamma_f=2.73\cdot 10^{13}$ s$^{-1}$ for Dy:CdO). The value for $\chi^{(3)}_{hot}$ measures the change of the free electron's effective mass as the medium is irradiated. The temperature of the free electron gas generally depends on the rate of absorption, and thus the fluence of the incident pulse. In turn, if temperature excursions are limited to a few thousand degrees Kelvin, the effective mass changes linearly with temperature, so that to first order the nonlinearity may be assumed to be a function of peak power density, as outlined in Eqs.(3-4). Therefore, $\chi^{(3)}_{hot}$ will depend on equilibrium plasma frequency, electron temperature, frequency-dependent conductivity, pulse duration, and will be of order $10^{-18}$ m$^2$/V$^2$ at the TH and $10^{-19}$ m$^2$/V$^2$ at FF.

The second order response is accounted for by evaluating Coulomb, magnetic, and convective contributions on the surface and volume occupied by the free electrons. In the pulsed regime, these nonlinear sources are described in [44, 45]. The second harmonic electromagnetic problem is also solved as outlined in Refs. [13, 46] using the finite element method. We thus analyze the impact of equations (3) and (4) and discuss the second order nonlinear effects separately.

The introduction of nonlocal and hot electron contributions alters significantly both the linear and nonlinear responses of the nanostructure. We begin our linear analysis by monitoring linear and nonlinear absorption of the multilayer stack. Results are shown in Fig. 5(a) and (b), respectively. Fig. 5(a) shows the linear pump absorption profiles calculated in the local and nonlocal approximations, for 60 fs and 120 fs incident pulses and low pump irradiance value (1 KW/cm$^2$) so as not to trigger nonlinear effects. Remarkably, the nonlocal terms, whose strength is predetermined by the Fermi energy and effective free electron mass, blueshift the resonance by nearly 200 nm. The effect of pulse duration may also be ascertained from the figure: the 120 fs pulse couples more strongly with the structures, leading to larger local fields, stronger light-matter coupling, and more pump absorption. In Fig.5(b) we show the effect of increasing input power density to 1 GW/cm$^2$, and simultaneously triggering hot electron contributions. We monitor pump absorption within the multilayer in the following scenarios: local, with (red line, square markers) and without hot electron contributions (black line, triangle markers); and nonlocal, with the hot electron nonlinearity included (blue line, circle markers) [see Fig. 5(b)]. The inclusion of the hot electron contribution in Fig.5(b) mitigates the blueshift displayed in the linear regime (Fig.5(a), broadening, redshifting, and smearing the resonance as the plasma frequency decreases (red line, square markers). The significance and complexity of this dynamical, time-dependent interplay cannot be overstated, and it impacts the fundamental and its harmonics. All absorption curves in Fig. 5 were calculated at $\vartheta_i=65°$.

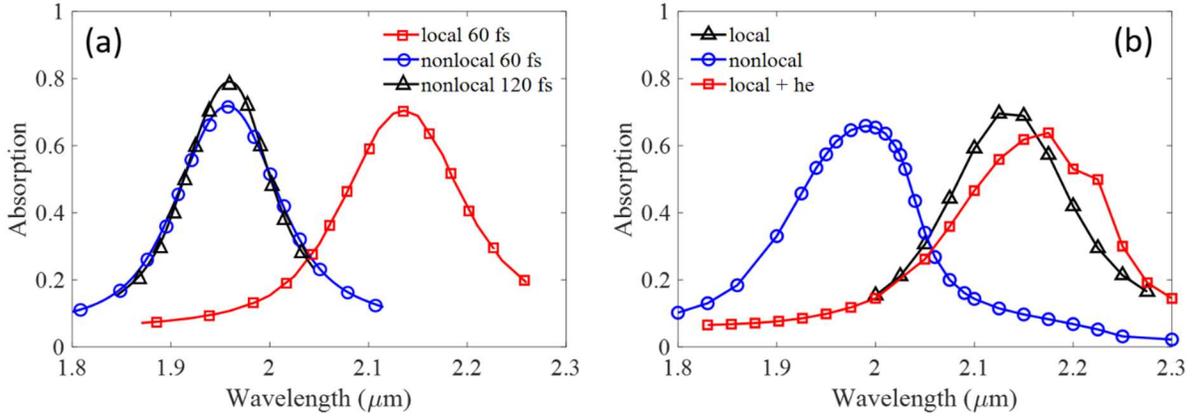

**Fig.5** (a) Pump absorption spectra calculated for low pump irradiance value (1 KW/cm$^2$) in the local (red line, square markers) and nonlocal (blue line, circle markers) approximation for a 60 fs incident pulse and nonlocal approximation for a 120 fs incident pulse (black line, triangle markers); (b) Pump absorption spectra calculated for high pump irradiance value (1 GW/cm$^2$) in the local (black line, triangle markers), nonlocal (blue line, circle markers) and local scenario with hot electron contributions (red line, square markers). All pulses are assumed 60fs long. Angle of incidence is $\vartheta_i=65°$ for all plots.

We then examine the average FIE calculated considering only nonlocal contributions [ $\chi^{(3)}_{hot,FF/TH}=0$ in equations (3) and (4)] and with the model that includes both nonlocal and hot electron contributions [equations (3) and (4) as shown], respectively [Fig. 6 (a) and (b)]. When comparing the maps in Fig. 6 with those in Fig.4 (a), two macroscopic differences may be discerned: (i) the nonlocality blue-shifts the structure' spectral response [47, 48]; (ii) FIE values drop for all wavelengths and angles of incidence. Moreover, while hot carrier contributions do not seem to alter the maximum FIE that may be achieved in the multilayer, significant distortion of both spectral and angular features are noticeable when hot electrons contributions are introduced [Fig. 6(b)]. The degree of distortion is proportional to pump irradiance value (not shown) [42].

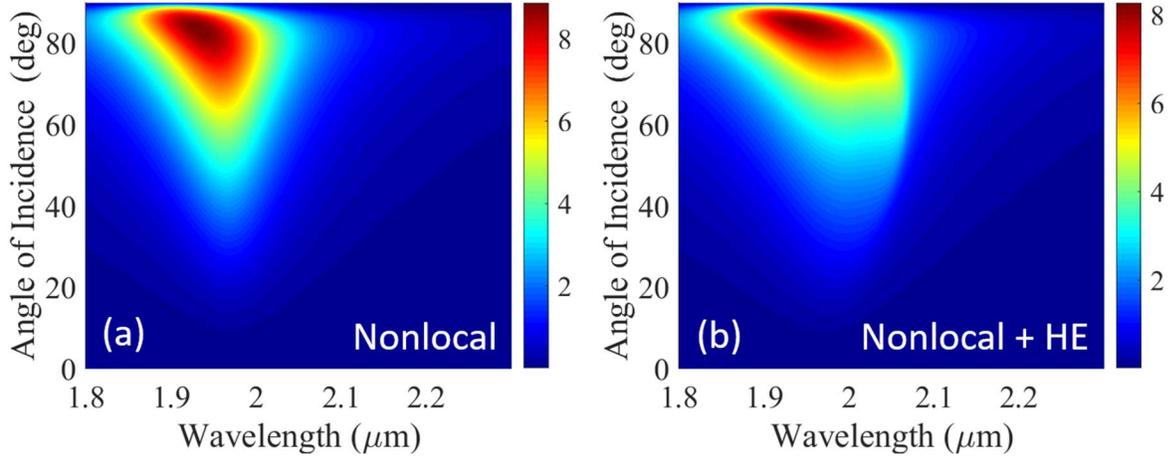

**Fig.6** FIE calculated as a function of wavelength and angle of incidence when (a) nonlocal contributions are present ( $\chi^{(3)}_{hot,FF/TH} = 0$ ) and (b) nonlocal scenario where hot carrier contributions are included - equations (3) and (4) implemented as shown. Fig. 5 (a) and (b) should be compared with Fig.4 (a): the nonlocality blueshifts the resonance, while the nonlinearity due to hot electrons imparts a redshift.

The impact of nonlocality and hot carriers on the third order nonlinear process are reported in Fig.7. We report a full comparison of TH conversion efficiency as a function of pump wavelength and angle of incidence in the local [Fig. 7 (a)], nonlocal – no hot electrons contributions [Fig. 7 (b)] and nonlocal – with hot electron contributions [Fig. 7 (c)] regimes. In Fig. 7 (d) we show three sections of the maps taken at the angle of incidence where maximum conversion efficiency is registered. The competing effects of nonlocal and hot carrier contributions are quite noticeable. By abandoning the local model to include nonlocal contributions [ $\chi^{(3)}_{hot,FF/TH} = 0$ , compare Fig. 7(a) and (b)] we record a significant blue-shift (~ 70 nm) of the third harmonic peak, and a slight shift in the angular response (maximum conversion efficiency moves from $\vartheta_i = 65°$ to $\vartheta_i = 69°$). Moreover, since the introduction of nonlocal effects hampers electric field enhancement [compare Fig. 4 (a) and Fig. 7 (a)], overall conversion efficiency decreases by one order of magnitude, to $10^{-8}$. On the other hand, the introduction of hot electron contributions [Fig.7 (c)] pushes the spectral resonance toward its original "local" spectral position, and causes distortions in the angular response that reflects what happens to the FIE [Fig. 6(b)]. However, while the FIE in the nonlocal scenario that includes the hot electron contributions [Fig. 6(b)] is comparable with the FIE in the absence of hot electron contributions [Fig. 6(a)], TH conversion efficiency is now of order $10^{-5}$, surpassing the simulated local and nonlocal scenarios [Fig. 7(a) and (b)]. One way to understand the large change in conversion efficiency is as follows: there are two sources of THG: the ENZ condition, which is resonant and dominates near the crossing point, and a non-resonant contribution due to bound electrons, which dominates far from the ENZ condition. Once the hot electron nonlinearity is triggered, newly generated TH photon flood the system [42].

Finally, in Fig. 7 (e) we show THG conversion efficiency for three different pulse durations, calculated using the hydrodynamic model discussed above. Longer pulses can resolve the resonance, with local fields reaching larger values and thus yielding improved conversion efficiencies and further resonance shifts. We note that the degree of distortion induced in the nonlocal scenario that includes hot electron contributions [Fig. 7(c) and green curve, triangle markers in Fig. 7(d)] strongly depends on pump irradiance and can lead to bistability at relatively low pump intensities. The onset of bistability is strongly suggested in Fig. 7(e), for 150fs incident pulses.

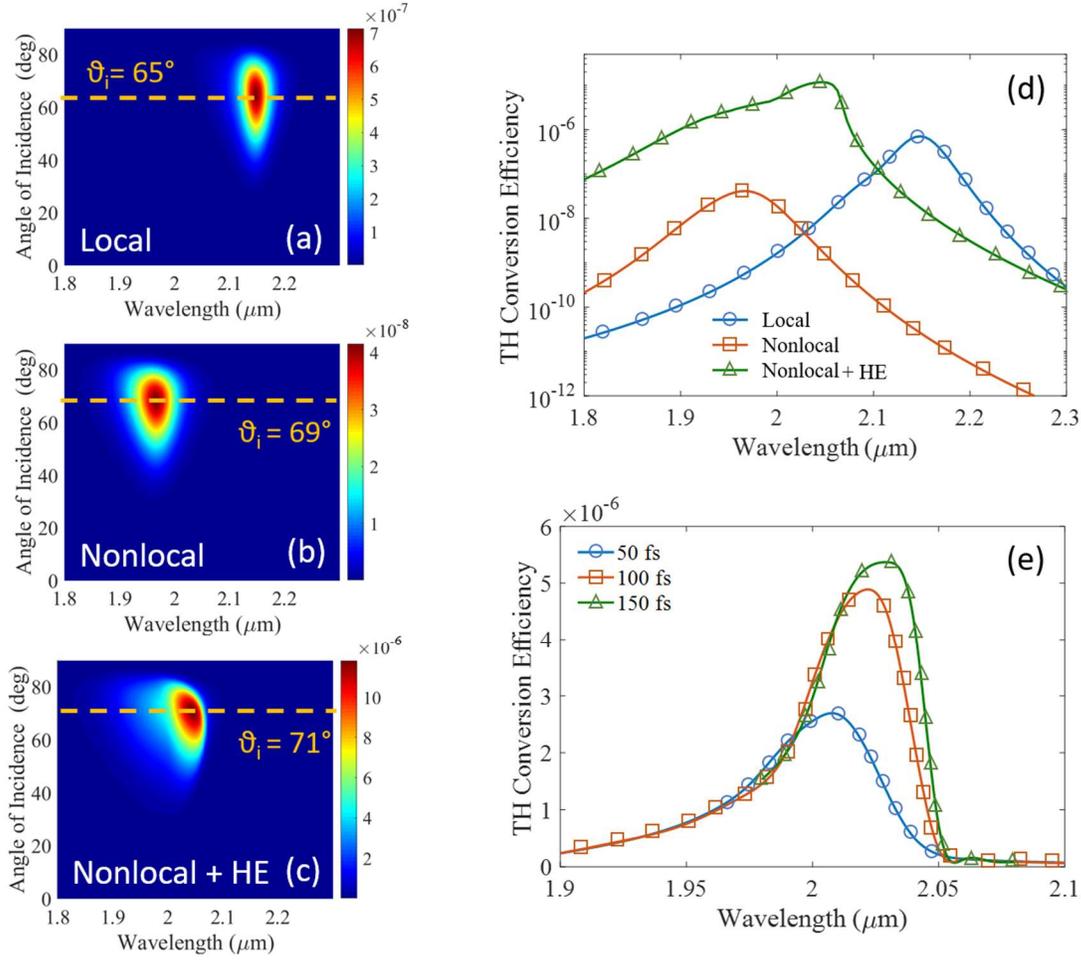

**Fig.7** TH conversion efficiency maps calculated as a function of pump wavelength and angle of incidence in a (a) local scenario [equations (1) and (2) are implemented], (b) nonlocal scenario [no hot carrier contributions are present, i.e. $\chi^{(3)}_{hot,FF/TH}=0$ in equations (3) and (4)] and (c) nonlocal scenario where hot carrier contributions are included - equations (3) and (4) implemented as shown. (d) sections of TH conversion efficiency maps assuming $\vartheta_i = 65°$ (blue line, circle markers - local scenario), $\vartheta_i = 69°$ (red line, square markers - nonlocal scenario) and $\vartheta_i = 71°$ (green line, triangle markers - nonlocal scenario assuming hot electrons contributions). (e) sections of THG conversion efficiency for $\vartheta_i = 65°$, for different pulse durations, as indicated.

As mentioned earlier, and as previously reported [38], the second order nonlinear response in these structures is significant. Additional effects occur at the metal/oxide interfaces due to the large field confinement and discontinuities in the free electron density [44, 49]. In Fig.8 we limit ourselves to depicting the predicted, reflected SHG conversion efficiency spectrum for $\vartheta_i = 65°$ incident angle, and different pulse durations (50 fs pulses – blue, solid line, 100 fs pulses – red, dashed line, 150 fs pulses, green, dotted line.) SHG undergoes shifts similar to THG as pulse duration is increased, but the maximum conversion efficiency is not altered significantly because SHG is more sensitive to field discontinuities than field localization. Finally, we compare the SH signal generated by the anisotropic, metal/conducting-oxide multilayer stack with the SH signal generated by a single, 20 nm-thick Dy:CdO layer set in a Kretschmann configuration [17], assuming a MgO prism and a 150 fs pulse incident at $\vartheta_i = 36°$ (black line, diamond markers). The figure suggests that at least one order of magnitude and 100nm separate the efficiency of the SH signals generated under the two different circumstances.

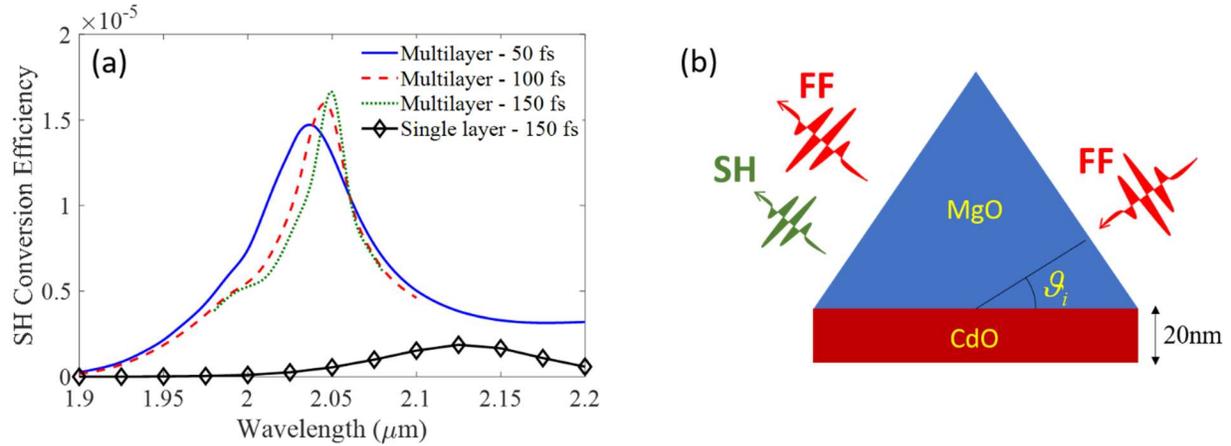

**Fig.8** (a) SHG conversion efficiency for the multilayer in Fig. 3 with $\vartheta_i = 65°$, for 50 fs pulse (blue, solid line), 100fs pulse (red, dashed line) and 150 fs (green, dotted line); (b) SHG conversion efficiency for a 20nm thick Dy:CdO layer in Kretschmann configuration (black line, diamond markers) assuming $\vartheta_i = 36°$ and 150 fs pulse.

**Conclusions**

In conclusion, we have attempted to clarify the role anisotropy plays in third order nonlinear processes. We demonstrated that a high degree of anisotropy can enhance nonlinear processes thanks to the high local field values that can be achieved. We also found that the sign of the anisotropy plays a different role in FIE and TH conversion efficiency when thicker slabs are investigated, suggesting that it may possible to exploit metals to achieve a large degree of anisotropy. Finally, we demonstrated that while an effective medium approach may be useful as a general guideline to design appropriate nanostructures, care should be exercised when non-homogeneous field distributions are achieved, as in the proposed multilayer. The introduction of nonlocal effects and hot electron contributions in fact reveals how these two effects compete and interact, leading not only to significant spectral and angular shifts, but also to the possible suppression/enhancement of the generated signal that depends on pump excitation conditions.

**Acknowledgment**

Research was sponsored by the Army Research Laboratory and was accomplished under Cooperative Agreement Number W911NF-20-2-0078.